\PassOptionsToPackage{unicode}{hyperref}
\PassOptionsToPackage{hyphens}{url}
\documentclass[
]{article}
\usepackage{lmodern}
\usepackage{amssymb,amsmath}
  \usepackage[T1]{fontenc}
  \usepackage[utf8]{inputenc}
  \usepackage{textcomp} 
\makeatletter
\@ifundefined{KOMAClassName}{
  \IfFileExists{parskip.sty}{%
    \usepackage{parskip}
  }{
    \setlength{\parindent}{0pt}
    \setlength{\parskip}{6pt plus 2pt minus 1pt}}
}{
  \KOMAoptions{parskip=half}}
\makeatother
\usepackage{xcolor}
\IfFileExists{bookmark.sty}{\usepackage{bookmark}}{\usepackage{hyperref}}
\usepackage{color}
\usepackage{fancyvrb}

\DefineVerbatimEnvironment{Highlighting}{Verbatim}{commandchars=\\\{\}}
\usepackage{framed}
\definecolor{shadecolor}{RGB}{248,248,248}
\newenvironment{Shaded}{\begin{snugshade}}{\end{snugshade}}

\newcommand{\CommentTok}[1]{\textcolor[rgb]{0.56,0.35,0.01}{\textit{#1}}}

\newcommand{\ControlFlowTok}[1]{\textcolor[rgb]{0.13,0.29,0.53}{\textbf{#1}}}
\newcommand{\DataTypeTok}[1]{\textcolor[rgb]{0.13,0.29,0.53}{#1}}
\newcommand{\DecValTok}[1]{\textcolor[rgb]{0.00,0.00,0.81}{#1}}

\newcommand{\FloatTok}[1]{\textcolor[rgb]{0.00,0.00,0.81}{#1}}

\newcommand{\KeywordTok}[1]{\textcolor[rgb]{0.13,0.29,0.53}{\textbf{#1}}}
\newcommand{\NormalTok}[1]{#1}
\newcommand{\OperatorTok}[1]{\textcolor[rgb]{0.81,0.36,0.00}{\textbf{#1}}}
\newcommand{\OtherTok}[1]{\textcolor[rgb]{0.56,0.35,0.01}{#1}}

\newcommand{\StringTok}[1]{\textcolor[rgb]{0.31,0.60,0.02}{#1}}

\usepackage{longtable,booktabs}
\usepackage{etoolbox}
\makeatletter
\patchcmd\longtable{\par}{\if@noskipsec\mbox{}\fi\par}{}{}
\makeatother
\usepackage{footnote}
\makesavenoteenv{longtable}
\usepackage{graphicx,grffile}
\def\fps@figure{htbp}
\makeatother
  \usepackage{fancyhdr}
  \usepackage{algorithm,algorithmic}
  \usepackage{bm}
  \usepackage[margin=1in]{geometry}
  \usepackage{colortbl}

\newcommand{\cF}{\mathcal{F}}
\newcommand{\cT}{\mathcal{T}}

\newcommand{\cX}{\mathcal{X}}

\newcommand{\cH}{\mathcal{H}}

\newcommand{\cL}{\mathcal{L}}
\newcommand{\cM}{\mathcal{M}}

\newcommand{\cD}{\mathcal{D}}

\newcommand{\cK}{\mathcal{C}} 
\newcommand{\CK}{\mathcal{C}}

\newcommand\myf{Q}

\newcommand\la{\lambda}

\newcommand\mY{\mathcal{Z}}

\newcommand{\rev}[1]{\textcolor{black}{#1}}

\newcommand{\bI}{\mathbf{I}}
\newcommand{\bK}{\mathbf{C}}

\newcommand{\vy}{\mathbf{y}}

\newcommand{\PP}{\mathbb{P}}

\newcommand{\FF}{\mathbb{F}}

\newcommand{\E}{\mathbb{E}}
\newcommand{\R}{\mathbb{R}}
\newcommand{\bx}{\bm{x}}

\newcommand{\vb}{\vartheta}

\DeclareMathOperator\argmax{arg\, max}
\DeclareMathOperator\argsup{arg\, sup}

\usepackage{booktabs}
\usepackage{longtable}
\usepackage{array}
\usepackage{multirow}
\usepackage{wrapfig}
\usepackage{float}
\usepackage{colortbl}
\usepackage{pdflscape}
\usepackage{tabu}
\usepackage{threeparttable}
\usepackage{threeparttablex}
\usepackage[normalem]{ulem}
\usepackage{makecell}
\usepackage[numbers]{natbib}
\title{Regression Monte Carlo for Impulse Control}
\author{Mike Ludkovski\footnote{Department of Statistics and Applied Probability, University of California, Santa Barbara, 93106-3110. Work partially supported by NSF DMS-1821240. I thank Zhuoli Jin for research assistance. \href{mailto:ludkovski@pstat.ucsb.edu}{\nolinkurl{ludkovski@pstat.ucsb.edu}}}}
\date{March 10, 2022}

\begin{document}
\maketitle
\begin{abstract}
I develop a numerical algorithm for stochastic impulse control in the spirit of Regression Monte Carlo for optimal stopping. The approach consists in generating statistical surrogates (aka functional approximators) for the continuation function. The surrogates are recursively trained by empirical regression over simulated state trajectories.
In parallel, the same surrogates are used to learn the intervention function characterizing the optimal impulse amounts. I discuss appropriate surrogate types for this task, as well as the choice of training sets. Case studies from forest rotation and irreversible investment illustrate the numerical scheme and highlight its flexibility and extensibility. Implementation in \texttt{R} is provided as a publicly available package posted on GitHub.
\end{abstract}

\section{Introduction}\label{introduction}

Stochastic impulse control is concerned with systems where the state process $(X_t)$ is subject to stochastic dynamics as well as repeated lumpy interventions or shocks by the controller. Such impulses make an instantaneous, rather than sustained, impact on $(X_t)$ and carry an instantaneous cost/reward. The goal of the controller is to maximize total expected (discounted) profit that is driven by the impulses and the continuous revenue function $\pi(X_t)$. Impulse control problems have a long history with manifold applications over the past 40+ years, see below. In particular, impulsive controls are common in commodities applications to describe management of natural resources or industrial capacity planning. Nevertheless, numerical methods for stochastic impulse control are surprisingly thinly studied, especially in comparison to the vast literature on numerical optimal stopping, and the emergent literature on numerical continuous control via Neural Networks.

In this article I propose to leverage  Monte Carlo based methods that have been developed for optimal stopping in order to create a related approach to stochastic impulse control. To do so, I rely on the interpretation of impulse control as a two-stage decision making; at each time instant, the controller must first decide whether to act or to wait; conditional on acting, in the second stage the controller picks the optimal action. This perspective reduces impulse control to repeated optimal stopping with an implicit payoff function specified via the so-called intervention operator $\cM$. In turn, it permits to import algorithms for multiple stopping problems after incorporating the computation of the intervention operator. Through this lens, solvers for impulse control can be built on top of related code for optimal stopping.

The proposed algorithm extends the realm of Regression Monte Carlo (RMC) methods to the setting of impulse control. It employs the main features of RMC --- statistical surrogates for a functional approximation of the continuation value and simulation for empirical training of these surrogates ---in the context of multiple impulsing actions. Additionally, I propose direct optimization of the surrogate over the action set to obtain optimal impulses. Methodologically, this strategy highlights the advantageous modularity of RMC which makes the paradigm applicable beyond classical stop/continue decisions. On the implementation side, the algorithm is coded in \texttt{R} and is integrated into the existing \texttt{mlOSP} "Machine Learning for Optimal Stopping Problems" package developed by the author over the past few years \cite{mlOSP}. Thus, \texttt{mlOSP} effectively subsumes numerical resolution of impulse control into the existing framework of RMC. The package is publicly available via GitHub at \url{github.com/mludkov/mlOSP} and offers reproducible vignettes. Thus, all the underlying code and case studies can be fully examined by the readers or other researchers, facilitating reproducibility and future extensions.

\subsection{Literature Review}
Relative to other types of stochastic control impulse control problems are rarely solvable explicitly, with only a few exceptions, see \cite{alvarez2004class,egami2008direct,christensen2014solution}. In part, this is because there are many problem ingredients to solve for: impulse thresholds, impulse targets, intervention function, value function, etc. Thus, typically only very special cases, such as time-stationary models with linear intervention costs and linear dynamics, have been studied in detail.

The most well known sub-case is when the state $(X_t)$ is one dimensional and is expected to be a renewal process when optimally controlled. The so-called
$(s,S)$ strategies focus on determining an impulse threshold $s$ and a target level $S$ and reduce computation of optimal strategy to a two-dimensional optimization over
the two constants $s,S$. $(s,S)$ strategies have been studied in Operations Research for over 20 years, formulations similar to the one I discuss  have appeared in optimal inventory \cite{bayraktar2010inventory,bensoussan2005optimality,hu2016s} and dividend payout problems \cite{azcue2019optimal,bayraktar2014optimal,czarna2011finetti}.

Another major application of stochastic impulse control has been in \emph{real options}, in the context of (ir)reversible investment \cite{aid2015explicit,alvarez2011optimal,Federico19}. Alternatively called the capacity expansion problem, this class of models considers gradual addition of capacity, e.g.~power generation capacity in the context of owning a fleet of electricity power plants. More sophisticated models\cite{bensoussan2019sequential,guthrie2012uncertainty}  treat separately the commodity price and the current capacity, leading to a two-dimensional impulse control formulation, with one exogenous and one endogenous component, see Section \ref{sec:capexpand}. A kind of a conceptual opposite to investment are \emph{harvesting} problems, especially the Faustmann problem of forest management \cite{alvarez2007optimal,alvarez2007taxation,belak2017general}.
The state variable  represents the current forest stand value; actions correspond to cutting trees down for sale, known as a "rotation". Thus impulses are down and are viewed as revenue rather than cost. I discuss the Faustmann problem further in Section \ref{sec:forest}. Other applied domains include control of foreign exchange rates by a central bank \cite{cadenillas2000classical} and management of energy retail prices \cite{basei2019optimal}.

The standard approach to numerical solution of impulse control is via quasi-variational inequalities, that reduce to a HJB-type partial differential equation. However, the non-local term in the equation that corresponds to the impulses makes numerical schemes quite challenging. So far there is limited literature available to handle it, \cite{azimzadeh2018convergence}. Indeed, eliminating this limitation of the HJB approach is one of the gaps I aim to address in the present publication that takes a completely probabilistic/statistical perspective and does not depend on the smoothness of the value function. Similarly, due to the associated analytic difficulties, the vast majority of works consider time-stationary one-dimensional impulse control on infinite horizon, see \cite{belak2017general} and \cite{elasri2020zero,elasri2020stochastic} for recent results on finite-horizon SIC. Multivariate SIC is treated in \cite{chen2013impulse,azcue2019optimal} among others.

The rest of the paper is organized as follows. Section \ref{sec:formal} introduces the problem and the new RMC-based algorithm. Section \ref{sec:workflow} discusses implementation, including how to evaluate optimal impulses. That Section concludes with a concrete example, illustrated with \texttt{R} code snippets.
Section \label{sec:results} contain two case studies, about forest rotation (Section \ref{sec:forest}) and two-dimensional capacity expansion (Section \ref{sec:capexpand}).

\section{Problem Formulation}\label{sec:formal}

To set the stage, we focus on diffusion models where the system state  \((X_t)\) in the absence of any impulses is assumed to satisfy a Stochastic Differential Equation of Îto type,
\begin{align}\label{eq:sde}
dX_t = \mu(X_t) \, dt + \sigma(X_t)\, dW_t,
\end{align}
where \((W_t)\) is a (multi-dimensional) Brownian motion and the drift \(\mu(\cdot)\) and volatility \(\sigma(\cdot)\) are smooth enough to yield a unique strong solution to \eqref{eq:sde}. The state space of $(X_t)$ is $\cX \subset \R$ and we assume the standard probabilistic structure of
\((\Omega, {\cal F}, ({\cal F}_t), \mathbb{P})\), where \(X_t\) is adapted to the filtration \(\mathbb{F}=({\cal F}_t)\).
Generalization to multiple dimensions is straightforward. 

The above is the uncontrolled dynamics, which are subject to control shocks.
To describe the latter, let \rev{$T < \infty$ be a given time horizon and denote by} $\mathfrak{A}$ the set of all admissible controls. An admissible control $A = \{ \tau_n, z_n\}$  is a double sequence such that
\begin{itemize}
  \item $\tau_n$ is an increasing sequence of $\FF$-stopping times, such that $\tau_n < \tau_{n+1}$ $\PP$-a.s. and $\lim_{n\to\infty} \tau_n = T$ $\PP$-a.s.

  \item $z_n$ is a sequence of random variables taking values in \rev{$\Xi \subset \mathbb{R}$} such that $z_n$ is $\cF_{\tau_n}$-measurable for every $n \ge 1$

  \item Integrability condition is satisfied $\sum_n \E[ e^{-r \tau_n}(1+z_n) ] < \infty$ which ensures finiteness of the discounted interventions.
\end{itemize}

A controlled process $X^{t,x,A}$ is indexed by its initial condition $X_t = x$ and the associated  admissible impulse strategy and satisfies for $s>t$
\begin{align}\label{eq:X-A}
  \rev{X^{t,x,A}_s} = x + \int_t^s \mu( X^{t,x,A}_r) dr + \int_t^s \sigma( X^{t,x,A}_r) dW_r + \rev{\sum_{n: t \le \tau_n \le s}} z_n.
\end{align}
The corresponding expectation conditional on $X_t = x$ is denoted by $\E_{t,x}[\cdot]$.

The dynamics in \eqref{eq:sde} then correspond to the case of no control: $A = \emptyset$, and one may view \eqref{eq:X-A} as the concatenation of the uncontrolled \eqref{eq:sde} on each $[\tau_{n}, \tau_{n+1})$ plus the instantaneous jumps $X^{t,x,A}_{\tau_n} = X^{t,x,A}_{\tau_n-} + z_n$.

Let $\pi: x \mapsto \mathbb{R}$ be the running reward function, and $\kappa: (x,z) \mapsto \R$ be the impulse cost function, representing the net revenue of applying impulse of size $z \in \R$ at time $t$ and state $x$. Typically $\kappa(x,z) < 0$, capturing the cost of moving $X_t$ from $x \in \cX$ to $x+z \in \cX$.  We assume a finite horizon $T$ with a respective terminal condition $\phi(X_T)$.
The controller aims to maximize discounted expected reward
\begin{align}\mathbb{E} \left[ \int_0^T e^{-r t}\pi(X^{0,x,A}_t) dt + \sum_{n : \tau_n < T} e^{-r \tau_n} \kappa( X^{0,x,A}_{\tau_n-}, z_n) +  e^{-r T)}\phi(X^{0,x,A}_T) \right] \rightarrow \max! \end{align}
on the horizon $T < \infty$, where $r \ge 0$ is the discount factor.  Above we assume that $\pi,\phi$ are such that $\E_{0,x}\left[ \int_0^T e^{-r t} |\pi(X^\emptyset_t)| dt + |\phi(X^\emptyset_T)| \right] < \infty$.

For an admissible strategy $A$, denote by
\begin{align}
  J_{t,T}(x;A) :=  \int_t^T e^{-r (s-t)}\pi(X^{t,x,A}_s) ds + \sum_{n : \tau_n < T} e^{-r (\tau_n-t)} \kappa(X^{t,x,A}_{\tau_n-}, z_n) + e^{-r \rev{(T-t)}} \phi(X^{t,x,A}_T).
\end{align}
the reward from applying the strategy $A$ on the interval $[t,T]$ and starting with $X_t = x$. Then our goal is to evaluate the \textbf{value function} \(V : [0,T] \times \cal{X} \to \mathbb{R}\),
\begin{align}
V(t,x) &:= \rev{\sup_{A \in \mathfrak{A}_t} \mathbb{E}_{t,x} \left[ \int_t^T e^{-r (s-t)}\pi(X^{t,x,A}_s) ds + \sum_n e^{-r (\tau_n-t)} \kappa(X^{t,x,A}_{\tau_n-},z_n) + e^{-r (T-t)}\phi(X^{t,x,A}_T) \right] } \\ \notag &=  \sup_{A \in \mathfrak{A}_t} \mathbb{E} \left[ J_{t,T}(x;A) \right],
\end{align}
\rev{where $\mathfrak{A}_t$ is the set of admissible strategies on $[t,T]$.}

The infinitesimal generator of the uncontrolled $X^{\emptyset}$  is
$$
\cL u(t,x) := r u(t,x) - \mu(x) \frac{\partial u(t,x)}{\partial x} - \frac{1}{2} \sigma^2(x) \frac{\partial^2 u}{\partial x^2} (t,x) - \frac{\partial u(t,x)}{\partial t}.
$$
Also define the intervention operator $$\cM u(t,x) := \sup_{z \in \Xi} \{u(t,x+z)\rev{+}\kappa(x,z)\}.$$ Optimality for the controller's actions implies that $V(t,x) \ge \cM V(t,x)$ for all $(t,x)$. At the same time, Ito's lemma implies that $\cL V(t,x) \ge \pi(x)$. Putting the two together yields the quasi-variational inequality (QVI)
\begin{align}\label{eq:qvi}
  \min \left( \cL V(t,x) - \pi(x), V(t,x)- \cM V(t,x) \right) = 0
\end{align}
with the boundary condition $V(T,x) = \phi(x)$. The analytic approach then characterizes $V$ as the viscosity solution of the QVI \eqref{eq:qvi}, see e.g.~Chapter 6 in Oksendal and Sulem \cite{oksendal2007applied}. HJB-driven methods either attempt to find a classical smooth solution to the QVI, or consider finite-difference schemes for \eqref{eq:qvi}; the challenge being the non-local operator $\cM$.

\subsection{Dynamic Programming Equation}\label{sec:dpp}
For the remainder of the article I adopt the discrete-time paradigm, where decisions are made at \(K\) pre-specified instances \(t_0=0 < t_1 < \ldots < t_k < t_{k+1} < \ldots < t_K = T\), where typically we have \(t_k = k\Delta t\) for a given discretization step \(\Delta t\). Henceforth, with a slight abuse of notation I index everything by \(k\) and work with \({\cal T} = (t_k)_{k=0}^K\), \rev{presuming that $t_k =k\Delta t$ for ease of exposition}. In particular this implies that we restrict $\tau_n \in \cT$ and rule out multiple instantaneous actions, so that $A$ consists of at most $K$ impulses.

The dynamic programming Bellman equation for impulse control on $[t,t+\Delta t]$ is:
\begin{multline}
  V(t,x)  = \max \Big( \E_{t,x}\left[ e^{-r \Delta t} V(t+\Delta t, X_{t+\Delta t}) + \int_t^{t+\Delta t} e^{-r (s-t)}\pi(X_s) ds  \right],  \\ \sup_{z \in \Xi} \E_{t,x+z} \left[ e^{-r \Delta t} V(t+\Delta t, X_{t+\rev{\Delta t} }) +\int_t^{t+\Delta t} e^{-r (s-t)}\pi(X_s) ds \right] + \kappa(x,z) \Big).
\end{multline}
Discretizing in time and writing $V(k,x) \equiv V(t_k,x)$, etc we substitute $\E_{t,x}[\int_t^{t+\Delta t} e^{-r(s-t)}\pi(X_s)ds] \simeq \pi(x)\Delta t$ and end up with
$$
V(k,x) = \pi(x)\Delta t + \max ( Q(k,x), M(k,x)),
$$
where $Q(k,x) =  \rev{\E_{k,x}}\left[ e^{-r \Delta t} V(k+1, X_{k+1})\right]$ is the so-called Q-value and $$M(k,x) = \cM Q(k,x) = \sup_{z \in \Xi} \{Q(k,x+z) + \kappa(x,z)\}.$$ The latter \emph{intervention operator} captures the value of making the best possible impulse.
In line with above, we view optimal impulse control as a two-stage sequential decision making. At each time period \(k\), the controller must decide whether to \rev{continue (no action) or act ($z \neq 0$)}. In the latter case, she must further select the best action $z^*$. This matches the appearance of $\max$ in $V(t,x)$---one should continue if the Q-value dominates the intervention value, and one should impulse otherwise. Within the Markovian structure of \eqref{eq:X-A} the impulse strategy can be encoded as
mapping each input $x$ according to the respective feedback \emph{action map} $\mY_k(x) \in \{ 0 \} \cup \Xi$:
\begin{align}\label{eq:Z-star}
\mY_k(x) = \arg\max_{z} \{ Q(k,x+z) + \kappa(x,z)\}  \cdot 1_{\{M(k,x) > Q(k,x) \}}.
\end{align}
The action map \(\mY_k\) gives rise to the action region
\[ {\cal S}_k := \{ x : \mY_k(x) \neq  0\} \subseteq {\cal X}, \]
where the optimal choice is to act.

Regression Monte Carlo proceeds by recursively constructing surrogates $\widehat{Q}(k,\cdot)$ that are used to induce the respective $\widehat{\mY}_k$ according to \eqref{eq:Z-star}. The inductive logical loop is achieved by employing $\widehat{\mY}_k$ to define the forward $J_{k,K}(x; \rev{\widehat{\mY}_{k:K}})$. To do so,
given any set of (admissible) action maps $\mY_{k:K}(\cdot)$ we define the corresponding discrete-time controlled state process $X^{x,\mY}$ according to the Euler scheme:  $X_{k} = x$ and
\begin{align}
  X^{x,\mY}_{k+1} = 
     X^{x,\mY}_k + \mu( X^{x,\mY}_k) \Delta t + \sigma (X^{x,\mY}_k) \Delta W_k + \mY_k  \qquad \text{where}\quad \mY_k \equiv \mY_k( X^{x,\mY}).
\end{align}
Note that the action is only applied at the end of the period. The respective total revenue along the path $X^{x,\mY}$ is then (setting $\kappa(x,0) \equiv 0$)
\begin{align}
  J_{k,K}(x; \mY_{k:K}) = \sum_{\ell=k}^{K-1} e^{-r(t_\ell-t_k)} \left\{\pi ( X^{x,\mY}_\ell)\cdot(t_{\ell+1}-t_\ell) + \kappa( X^{x,\mY}_\ell, \mY_\ell) \right\} + e^{-r(T-t_k)}\phi(X^{x,\mY}_K).
\end{align}
On a given path, we can also record the pathwise realized impulse times $\tau_n$ and respective impulses $z_n$:
\begin{align}\label{eq:pathwise-tau}
  \tau_n & = \min \{ k > \tau_{n-1} : \rev{{\mY}_k (X^{x,{\mY}}_k) > 0} \}  \\
  z_n &= \rev{{\mY}_k (X^{x,{\mY}}_{\tau_n})}. \label{eq:pathwise-impulse}
\end{align}

Denoting by $\mY^*$ the optimal action map, we have that the true Q-value satisfies
\begin{align}\label{eq:path-reward}
 Q(k,x) = \E_{k,x} \left[ J_{k,K}(x; \mY^*_{k:K}) \right].
\end{align}
In RMC, $Q(k,\cdot)$'s are replaced with $\widehat{Q}(k,\cdot)$; the latter induce $\widehat{\mY}_k$. Finally, $\widehat{\mY}_{k:K}$ is used via \eqref{eq:path-reward} to characterize and fit $\widehat{Q}(k-1,\cdot)$. The resulting loop is initialized with \(\widehat{V}(K,x)=\phi(x)\) and proceeds as follows:

For \(k=K-1,\ldots,1\) repeat:

\begin{enumerate}
\def\labelenumi{\roman{enumi})}
\item
  Learn the Q-value \(\widehat{Q}(k,\cdot) \simeq {\E}\left[ e^{-r \Delta t}\widehat{V}(k+1, X_{k+1}) \big|\, X_k = \cdot \right] \);
\item Evaluate the intervention function $\widehat{M}(k,\cdot) = \sup_{ z\in\Xi} \{\widehat{Q}(k,\cdot+z) + \kappa(\cdot,z)\}$
\item Set
 \begin{align}\label{eq:hat-Z}
 \widehat{\mY}_k(x) := \left\{ \begin{aligned} 0 & \quad\text{ if } \widehat{Q}(k,x) > \widehat{M}(k,x)  \\
 \argmax_z \{\widehat{Q}(k,x+z) + \kappa(x,z)\} & \quad\text{ otherwise.} \end{aligned} \right.\end{align}

\item Record $\widehat{V}(k,x) := \max \bigl( \widehat{Q}(k,x), \widehat{M}(k,x) \bigr) + \pi(x) \Delta t$.
\end{enumerate}

Note that in principle the entire $\widehat{V}$ is superfluous. Indeed, we have generically that for any $w$
$$ Q(k,x) = \E\left[ J_{k,k+w}(x; {\mY}^*) + e^{-r \Delta t w} V(k+w, X^{k,\mY^*}_{k+w}) \right].$$
The look-ahead horizon \(w \in\{ 1,\ldots, K-k\}\) allows to combines pathwise rewards based on $\widehat{\mY}$ and the approximate value function \(\widehat{V}\) \(w\)-steps into the future \citep{EgloffKohlerTodorovic07}.
We focus on the cases $w=1, w=K-k$ and $w$ fixed that correspond to learning
\begin{align*}
w=1: \qquad \tilde{Q}^{(1)}(k,x) & = \E_{k,x} \left[ \pi(x) \Delta t +  e^{-r \Delta t} \widehat{V}(k+1, X^{k,\emptyset}_{k+1}) \right] \\
w \text{  fixed}: \qquad \tilde{Q}^{(w)}(k,x) & = \E_{k,x} \left[ J_{k,k+w}(x; \widehat{\mY}_{k:k+w}) + e^{-r w \Delta t}\widehat{V}(k+w, X^{k,\widehat{\mY}}_{k+w}) \right] \\
w=K-k: \qquad \tilde{Q}^{(LS)}(k,x) & = \E_{k,x} \left[ J_{k,K}(x; \widehat{\mY}_{k:K}) + e^{-r (K-k)\Delta t} \phi(X^{k,\widehat{\mY}}_{K}) \right]
\end{align*}
The choice $w=1$ is analogous to the Tsitsiklis-van Roy \citep{TsitsiklisVanRoy} scheme for optimal stopping:  simulate one-step-ahead paths and regress
$\pi(X_k)\Delta t + \widehat{V}(k+1,X_{k+1})$ against $X_k$.  The choice $w=K-k$ is analogous to the Longstaff-Schwartz \citep{LS} scheme, where we regress the full future rewards to go on the interval $\{k,k+1,\ldots,K\}$  against \(X_k\).
These choices are not numerically identical, because \(\widehat{Q}(k+1, x) \neq \E\left[ J_{k+1,\ell}(x; \widehat{\mY}_{k:\ell}) |\, X_{k+1}=x \right]\) for different $\ell$'s due to the approximation error. In all examples below, we utilize the Longstaff-Schwartz version with $w=K-k$, so that $\widehat{V}$ is never computed until the very end.

\subsection{Algorithm}

The proposed RMC approach reduces optimal impulse control to a double sequence of probabilistic function approximation tasks.
The primary task entails fitting a functional approximator $\widehat{Q}$ based on empirical simulations and then utilizing a statistical model to capture the observed input-output relationship.
 To do so, we define a regression model and the training set used as input to the regression model. The three ingredients are the inputs \(x^{1:N} \in \cal{X}\), the outputs $y^{1:N}$ and the approximation class $\cH_k$. The inputs are the sampled states at step \(k\). The outputs are viewed as a random realization $Y(x)$ of the pathwise reward starting at \((k,x)\) such that \(\mathbb{E}_{k,x}[Y(x)] = Q(k,x)\). Specifically, they are the realizations of $J_{k,K}(x; \widehat{\mY}_{k:K})$ along a set of independent paths $x^{(k),n}$ that start with $x^{(k),n}_k = x^n$.
 Statistically these $y^{1:N}$ are linked to the inputs by the observation model
\begin{align}\label{eq:stat-model}
Y(x) = \E[Y(x)] + \epsilon(x), \qquad \E[ \epsilon(x)] = 0, \quad\mathbb{V}ar(\epsilon(x)) = \sigma^2_\epsilon(x).
\end{align}
Given  a training collection \( \cD = x^{1:N}\) henceforth called the simulation design, we collect the simulation outputs \(y^{1:N} = Y(x^{1:N})\) and then obtain the approximate continuation value $\widehat{Q}(k,\cdot)$ (viewed as a statistical object, rather than say a vector of numbers) as the empirical $L^2$ minimizer in the given function space \({\cal H}\). Namely we minimize the penalized mean squared error from the observations,
\begin{align}\label{eq:spline}
\widehat{Q}(k,\cdot) = \arg \min_{f \in {\cH}} \sum_{n=1}^{N} (f(x^n) - y^n)^2 + \la \| f \|_{\cH}.
\end{align}
The summation in \eqref{eq:spline} is a measure of closeness of $f$ to data, while the right-most term penalizes the fluctuations of $f$ to avoid over-fitting.

Indexing everything by the time steps \(k=1,\ldots,K\) and allowing for time dependence we summarize the following notation:
\begin{itemize}
\item
  \(N_k\): number of training inputs at step \(k\);
\item
  \({\cal D}_k\): simulation design, i.e.~the collection of training inputs \(x^{1:N_k}\), \(|{\cal D}_k| = N_k\)
\item
  \({\cal H}_k\): functional approximation space where \(\widehat{Q}(k,\cdot)\) is searched within;
\item
  \(y^{1:N_k}_k\) pathwise samples of reward-to-go used as the responses in the regression model.
\end{itemize}

Equipped with above, Algorithm \ref{alg:1} presents the overall scheme that abstracts from the regression module for fitting \(\widehat{Q}(k,\cdot)\) and subsequently $\widehat{M}(k,\cdot)$. The algorithm matches the \textbf{mlOSP} template from \cite{ludkovski2020mlosp} as implemented in the eponymous \texttt{R} package.

\begin{algorithm}
\begin{algorithmic}[1]
\REQUIRE{ $K=T/\Delta t$ (time steps), $(N_k)$ (simulation budget per step)}, $w$ (path lookahead)
\FOR{$ k=K - 1, \ldots, 0$ }
\STATE Generate training design $\mathcal{D}_{k} := (x^{(k),1:N_k}_k)$ of size $N_{k}$
\STATE Set $y^{1:N_k}_{k+1} \leftarrow 0$    // pathwise rewards
\FOR{$ \ell=k+ 1, \ldots, k+w \wedge K$ }
\STATE Sample $x^{(k),n}_{\ell-1} \mapsto x^{(k),n}_{\ell} \quad $  // pathwise controlled trajectories
\STATE Set $y^n_{k+1} \leftarrow y^n_{k+1} + \pi( x^{(k),n}_{\ell-1}) \Delta t$
\STATE Evaluate $m^{(k),n}_\ell = \widehat{M}(\ell,x^{(k),n}_{\ell})$ and $q^{(k),n}_\ell = \widehat{Q}(\ell,x^{(k),n}_{\ell})$
\STATE Set $x^{(k),n}_\ell \leftarrow x^{(k),n}_\ell + \widehat{\mY}_\ell(x^{(k),n}_{\ell}) \qquad\quad$          for $n$ where $m^{(k),n}_\ell > q^{(k),n}_\ell$  // impulse
\STATE Set $y^n_{k+1} \leftarrow y^n_{k+1} + \kappa(x^{(k),n}_\ell, \widehat{\mY}_\ell(x^{(k),n}_{\ell}) ) $ for $n$ where $m^{(k),n}_\ell > q^{(k),n}_\ell$
\ENDFOR
\STATE Set $ y^n_{k+1} \leftarrow y^n_{k+1} +
e^{-r w \Delta t} \max( \widehat{Q}\bigl(k+w, x^{(k),n}_{k+w} \bigr), \widehat{M}\bigl(k+w, x^{(k),n}_{k+w} \bigr))$.
\STATE Fit $\widehat{Q}(k,\cdot)$ by regressing $\{y^{1:N_k}_{k+1}\}$ on $\{x^{(k),N_k}_k\}$
\ENDFOR
\STATE Return fitted objects $\{ \widehat{Q}(k,\cdot) \}_{k=0}^{K-1}$
\end{algorithmic}
\caption{Regression Monte Carlo for Impulse Control based on \textbf{mlOSP} template. \label{alg:1}}
\end{algorithm}

The output of Algorithm \ref{alg:1} is the approximate action maps \(\widehat{\mY}_k(\cdot)\). Once computed, they induce the expected reward
$\E\left[ J_{0,K} (x; {\widehat{\mY}_{0:K}}) \big|\, X_0 = x\right].$
which can be evaluated over an out-of-sample set of test scenarios. Thus, we compute the sample average reward across a fresh set of \(x^{1:N',\widehat{\mY}}_k, k=1,\ldots,K\), \(x^{n'}_0 = x)\),
\begin{align}\label{eq:tildeV}
\check{V}(0,x) = \frac{1}{N'} \sum_{n'=1}^{N'} \left\{ \sum_{k=0}^{K-1} e^{-r t_k} \pi( x^{n',\widehat{\mY}}_k) (t_{k+1}-t_k) \rev{+} \sum_{m: \tau^{n'}_m < T} e^{-r \tau^{n'}_m}\kappa( x^{n',\widehat{\mY}}_{\tau^{n'}_m}, z^{n'}_m) \right\}
\end{align}
where $(\tau_m,z_m)$ are the pathwise impulse times and impulse amounts, see \eqref{eq:pathwise-tau}-\eqref{eq:pathwise-impulse}.  Note that $\check{V}(0,x)$ is an unbiased estimator of $\E\left[ J_{0,K} (x; {\widehat{\mY}_{0:K}}) \big|\, X_0 = x\right]$ and the latter is a lower bound on the true optimal expected reward, so that $$\E[ \check{V}(0,x)] < V(0,x).$$

\subsection{Relation to Stationary Impulse Control}

The cited analytical works consider the infinite horizon case of solving for  $$v(x) = \E_x\left[ \int_0^\infty e^{-rs}\pi(X_s) ds + \sum_n e^{-r \tau_n} \kappa(X_{\tau_n-},z_n) \right].$$ Assuming time-stationary dynamics for $X^\emptyset$, the optimal strategy is also time-stationary, meaning that there is a feedback action map $\mY^*(x)$ that is independent of $t$ and fully characterizes the impulses.

In contrast, the solution constructed above is explicitly time-dependent, as it is specified by $\widehat{Q}_k$ that are intrinsically distinct for different $k$.
Nevertheless, when far from the horizon $K$, the time-dependence should be intuitively weak, and we expect to recover the time-stationary $\mY^*(x)$\rev{.}
Indeed, informally if we parameterize in terms of time-to-maturity $\tilde{V}_{K-k}(\cdot) := V(k, \cdot ; K)$ we observe that by induction, $\widehat{V}(k,\cdot; K) = \widehat{V}(k+1, \cdot; K+1)$ since they both correspond to running the backward RMC algorithm for $K+1-(k+1) = K-k$ rounds. Thus, $\tilde{V}_\ell(\cdot)$ is well defined and the regime $\ell \to \infty$ corresponds to being far from the terminal condition so that we expect $\tilde{V}_\ell \to v(\cdot)$ as $\ell \to \infty$.

The above perspective suggests that for $k$ small, we can validate our $\widehat{Q}(k,\cdot)$ by comparing with $v(\cdot)$. Conversely, we may approximate $v(\cdot)$ by $\widehat{Q}(k,\cdot; K)$ for $k$ small and $K$ large. To speed up that convergence, we recall \emph{model predictive control} where during forward scenario generation one uses $\mY_k$ (rather than $\mY_\ell, \ell=k+1,\ldots, k+w$ for all time-steps. In other words, we compute expected reward based on a time-stationary control that is derived from the latest  (in the sense of backward induction) surrogate $\widehat{Q}(k,\cdot)$. The resulting $J_{k,K}(x; \widehat{\mY}_k)$ captures the reward over $K-k$ steps and its expectation would be a good approximation of $v(x)$ for $K$ large. Using model predictive control in Algorithm \ref{alg:1} is analogous to  a policy iteration search for infinite-horizon problems; it reinterprets $K$ as the receding horizon depth.

\section{Implementation}\label{sec:workflow}
In general, there is a huge range of potential statistical models for empirically fitting a $\widehat{Q}(k,\cdot)$. Thus, any statistical learning framework could be applied; see the \texttt{mlOSP} package that allows the use of more than a dozen different regression modules, linking to the vast library of \texttt{R} regression packages, from random forests to support vector machines. However, the
 fact that $\widehat{Q}$ is necessary to evaluate $\widehat{M}$ imposes requirements on what would be \emph{good} surrogates. For example, piecewise models (like a random forest, multivariate adaptive regression splines (MARS), or a hierarchical linear model) would tend to be inappropriate, as they would lead to discontinuities in defining $\widehat{\mY}_k(x)$ and hence unstable schemes due to error backpropagation. Similarly, polynomial bases might be problematic since their gradient tends to be highly oscillatory and therefore lead to unstable behavior in $\widehat{\mY}_k(x)$. Overall, we seek \emph{smooth} regression models with an interpretable gradient.

 Our two main proposals are \emph{smoothing splines} (SS) and Gaussian processes (GP). Splines intrinsically target $C^2$ fits, and tend to be highly robust to noisy data. Their main limitation is poor scalability, but otherwise they are a great default choice in 1 or 2 dimensions. Gaussian Processes yield smooth functional interpolators that work well with non-uniform training designs. Moreover, despite being non-parametric,  GPs yield analytic gradients.

Both SS and GPs consider \eqref{eq:spline} for a certain smoothing parameter $\la \ge 0$ and a
Reproducing Kernel Hilbert Space (RKHS) $\cH$.
The representer theorem implies that the minimizer of \eqref{eq:spline} therefore has an expansion in terms of the RKHS eigen-functions
\begin{equation}\label{eq:reg-regression}
\widehat{Q}(\cdot) = \sum_{n=1}^{N} \alpha_n \cK(\cdot,{x}^{n}) \rev{+ \sum_j \beta_j N_j(\cdot)},
\end{equation}
\rev{where $N_j$ span the null space of $\cH$.}
Note that this is a non-parametric fit since it involves the sum over the data-driven $\cK(\cdot, x^n), n=1,\ldots,N$.

\textbf{Smoothing splines:}  Thin-plate splines take the RKHS $\cH_{TPS}$ as the set of all twice continuously-differentiable functions with
 $\| f\|^2_{\cH_{TPS}} = \int_\R \{f''(x)\}^2 dx$.
  As $\la \to \infty$,  the optimization in \eqref{eq:spline} penalizes any convexity and ultimately reduces to the linear fit $\widehat{\myf}({x}) = \beta_0 + \beta_1 {x}$.
  \rev{Indeed, the null space of $\cH_{TPS}$ consists of affine functions, cf.~the second term in \eqref{eq:reg-regression}.}
   A common parametrization for the smoothing parameter $\la$ is through the effective degrees of freedom statistic df$_\la$; one may also select $\la$ adaptively via cross-validation or \rev{Maximum Likelihood Estimation (MLE)} \cite[Chapter 5]{FHT}. 
The respective eigenfuctions are $\cK_{TPS}(x, x') = | x - x'|^2 \log | x - x' |, $ and
 optimization of \eqref{eq:spline} gives a smooth $\mathcal{C}^2$ surrogate that 
has the explicit form
\begin{equation}\label{eq:tps-solution}
\widehat{\myf}_{TPS}(x) = \beta_0 + \beta_1 x + \sum_{n=1}^{N} \alpha_n |x-x^{n}|^2 \log |x-x^{n}|.
\end{equation}
See \cite{Kohler08spline} for implementation of RMC via  splines.

\textbf{Gaussian Processes:} GPs start with a positive definite kernel $c(x,x')$ which defines the function space $\cH_{\cK}$ and take $\lambda = 1/2$. The corresponding norm $\| \cdot \|_{\cH}$ has a spectral decomposition in terms of differential operators \cite[Ch.~6.2]{WilliamsRasmussenBook}. An intuitive interpretation is that GPs find $\widehat{Q}(\cdot)$ through applying Gaussian conditioning equations to the training data $(x^{1:N}, y^{1:N})$. To do so, a \rev{GP regression (GPR)} model specifies the covariance function $c(x,x')$ and a mean function $m(x)$, assumed for simplicity to be constant $m(x) \equiv \beta_0$. The GP estimate is then
\begin{align}\label{eq:gp_mean}
\widehat{\myf}_{GP}({x}) = \beta_0 + \rev{\CK(x)}^T(\bK + \sigma^2_\epsilon \bI)^{-1}(\vy-\beta \bm{1} )
\end{align}
where $\bI$ is the $N\times N$ identity matrix, $\bm{1}$ is the $N$ vector of 1's,
\begin{align}
  & \vy = [y^1,\ldots,y^{N} ]^T, \qquad \rev{\CK(x)^T = [ c(x, x^1; \vb), \ldots, c(x,x^{N};  \vb)]},
\end{align}
and $\bK$ is $N \times N$ covariance matrix described through the kernel function $\bK_{i,j}=c(x^i,x^j;\vb)$. 
The \rev{parameter} $\sigma^2_\epsilon$  comes from the observation noise in \eqref{eq:stat-model}, interpreted as being i.i.d.~Gaussian with the respective variance, \rev{ and is to be inferred with the rest of the GP hyperparameters.}

A GPR \rev{is implemented} by fitting the hyper-parameters $\vartheta$ governing the covariance kernel, the mean function and the observation noise. The user first specifies a parametric family and then optimizes, typically through the nonlinear MLE procedure.
The GP kernel $c(x,x')$ controls the smoothness (in the sense of differentiability) of \rev{$\widehat{Q}_{GP}$} and hence the roughness of its gradient.
 A popular choice for  $c(\cdot,\cdot)$ is the (anisotropic) squared exponential (SE) family, parametrized by the lengthscale $\ell_{\mathrm{len}} $ and the process variance $\sigma_p^2$ :
\begin{equation}
 c_{SE}(x,x'): = \sigma_p^2\exp{\Big(-  \frac{(x - x' )^2  }{2 \ell^2_{\mathrm{len}}}\Big)}.\label{eqn:se-kernel}
\end{equation}
The SE kernel~\eqref{eqn:se-kernel} yields infinitely differentiable fits $\widehat{Q}(k,\cdot)$ and has hyperparameters $\vb := ( \ell_{\mathrm{len}},  \sigma^2_{p}, \sigma^2_{\epsilon})$. Other popular kernels include those from the Mat\'ern family.

 \emph{Remark:} Artificial Neural Networks (ANNs) with smooth activation functions could be another appropriate framework, facilitating training via back-propagation. \rev{Machine learning libraries like TensorFlow provide facilities for fitting ANNs, as well as efficiently differentiating them in order to evaluate $\partial_x \widehat{Q}$ as in the next section. Those libraries are native to Python, rather than R, and are not yet supported in the current version of \texttt{mlOSP}.}

\subsection{Approximating the Intervention Function}

The computation of $\widehat{\mY}_k(x)$ is embedded deep in Algorithm \ref{alg:1} and drives the outputs $y^{1:N}_{k+1}$ used to fit $\widehat{Q}$. In this section I discuss how that piece of the solver should be implemented. The base implementation is to directly solve \eqref{eq:hat-Z} by calling an optimization sub-routine. The objective function is given \emph{implicitly} in terms of the object $\widehat{Q}(k,\cdot)$ so ostensibly a general-purpose, gradient-free optimizer may be needed. Given that $\widehat{M}(k,\cdot)$ has to be evaluated repeatedly on each forward path emanating from each training input $x^{(k),n}$ this is the major computational bottleneck. To overcome it, several efficiencies could be exploited.

First, one may speed up the computation by using a gradient-based optimizer. This requires to specify not just $\widehat{Q}(k,\cdot)$ but also $\partial_x \widehat{Q}(k,\cdot)$ in an explicit functional way. The latter is available for several types of surrogates, including splines and GPs. For the latter,  we recall that given a fitted GP model $\widehat{Q}(k,\cdot)$, its gradient  forms another GP with the respective mean at input $x_*$ being 
\begin{align}\label{eq:grad-mean}
 &g_*(x_*) := \frac{\partial \widehat{Q}}{\partial x}(x_*) = \frac{\partial c}{\partial x}(x_*, {\bx}) (\bK + \sigma^2_\epsilon \bI)^{-1}(\vy-\beta_0 \bm{1}), 
\end{align}
Thus, the gradient of the surrogate is $g_*(x_*)$ in \eqref{eq:grad-mean} which can be interpreted as formally differentiating the expression in \eqref{eq:gp_mean} with respect to $x$.
For example, for the SE kernel \eqref{eqn:se-kernel} we have:
\begin{align}\notag
&\frac{\partial c_{SE}}{\partial x}(x,x') = \frac{x' - x  }{\ell^2_{\mathrm{len}}} c_{SE}(x,x').
\end{align}

Second, one may exploit specific features of the problem setting. As a foremost example, I now discuss the common case where $\kappa(x,z)$ is linear in $z$, namely $\kappa(x,z) = c_0 z + c_1$ for some constants $c_0,c_1$. In this situation, the optimization defining $\widehat{M}(k,x)$ simplifies considerably. Indeed, the first order conditions reduce to searching for the ``global'' impulse target $S^*_k$:
\begin{align}\notag
 S^*_k & := \argsup_z \{ \widehat{Q}(k,x+z) + \kappa(x,z) \} = \argsup_y \{ \widehat{Q}(k,y) +c_0(y-x) + c_1 \} \\
 \Longleftrightarrow \partial_x \widehat{Q}(k,S^*_k) &= -c_0.
\end{align}
In particular, the target level $S^*_k$ is independent of the current state $x$, and moreover can be determined by a single root search on the gradient of the value function. This drastically simplifies and stabilizes the numerics, since we just need to determine $S^*_k$ once, and can then immediately compute $\widehat{M}(k,x) = \widehat{Q}(k,S^*_k)+c_0 S^*_k - c_0 x + c_1$ for any $x$. Consequently, the action region is $\widehat{\mathfrak{S}}_k  =\{ x : \widehat{Q}(k,S^*_k) - \widehat{Q}(k,x) > c_0(x-S^*_k) - c_1 \}$.

Third, when computing $\widehat{M}(k,x^{1:N_k})$ for each training input $x^{1:N_k}$, one can record and save the resulting optimal impulse amount $z^{1:N_k}_k$. Then in subsequent calls, instead of again solving for $\widehat{M}(k,x')$ at some new $x'$ by re-rerunning the optimizer, one may instead train
 a separate independent functional representation $\hat{Z}(k,\cdot)$ based on the dataset $(x^{1:N_k}_k, z^{1:N_k}_k)$. Thus, we could build an auxiliary surrogate $\hat{Z}(k,\cdot)$ (e.g.~through another GP surrogate) and then use $\hat{Z}(k,x')$ instead of $\arg\max_z \{ \widehat{Q}(k,x'+z) + \kappa(x',z) \}$. This substitutes the prediction $\hat{Z}(k,\cdot)$, which is typically much faster to compute, instead of calling the optimizer.

\subsection{Training Designs}\label{training-designs}

To train the regression surrogate, the user must supply the simulation design(s) \({\cal D}_k\). See \cite{mlOSP} for a detailed description of various options for training optimal stopping emulators within \texttt{mlOSP}. With impulse control, the major difference is that $(X_k)$ is no longer autonomous. Thus, it no longer makes sense to construct $\cD_k \sim p(X_k)$ as a sample from the uncontrolled dynamics. Instead, I propose to directly specify  ${\cD_k}$, building on the idea that the quality of $\widehat{Q}$ reflects the geometry of ${\cD_k}$ ---one learns best in regions where the training samples lie. Since the optimally controlled $(X^*_k)$ typically has a stationary distribution (modulo time-dependence imposed by the finite horizon), one may select a training \emph{region} based on a prior guess of the latter. For example, one may choose a hyper-rectangle $\bar{\cD}$ and then set $\cD_k$ as a finite, space-filling sample from $\bar{\cD}$,  yielding $\cD_k$ that looks like a sample from a uniform density on $\bar{\cD}$. Some of the ways to achieve this are This can be achieved either through a deterministic lattice, or i.i.d.~Uniform (stratified) sampling, or a low-discrepancy (Quasi Monte Carlo) sequence. All these choices will
\begin{itemize}
  \item Direct specification of $\cD_k$, e.g. as a fixed lattice \texttt{seq(a,b,by=$\Delta x$)};

  \item Probabilistic sampling of $\cD_k$, either using i.i.d. Uniforms, or a variance reduced variant of the former (e.g.~stratified sampling) or a Latin Hypercube sampling method (package \texttt{lhs} in R);

  \item Generation of $\cD_k$ from a (scrambled) low-discrepancy sequence (LDS), such as Sobol (package \texttt{randtoolbox} in R). Note that in this case $\cD_k$ is deterministic. This is specified by the \texttt{qmc.method} field that supports LHS (default) and various LDS.
\end{itemize}

Beyond targeting a uniform density of training samples on a given region, one may also take non-uniform $\cD_k$ that preferentially place more training inputs in some parts of $\cX$. For example, we may put more $x^k$'s in the region where we expect the impulse target to be, in order to improve the quality of $\widehat{Q}$ there, and hence the quality of $\widehat{M}$. The underlying intuition is that learning is achieved through exploration (sampling a diverse collection of $x^k$'s) and exploitation (sampling $x^k$'s that are likely to be encountered on forward controlled paths).

A further training option that I highlight is  \textbf{replication.} A replicated design is akin to a Monte Carlo forest, in the sense that some training inputs appear multiple times. In a most common \emph{batched} design, we have \(N_{unique}\) distinct sites, the so-called macro-design, and each unique \(x^n\) is then repeated \(N_{rep}\) times, so that
\begin{align}\label{eq:rep-design}
 {\cal D} = \{ \underbrace{x^{1}, x^{1}, \ldots, x^{1}}_{N_{rep} \text{ times}}, \underbrace{x^{2}, \ldots}_{N_{rep} \text{ times}}, x^{3}, \ldots, \ldots, x^{N_{unique}} \},
\end{align}
where the superscripts now index \emph{unique} inputs and the total training budget for $\widehat{Q}(k,\cdot)$ is \(|\cD| = N_{unique} \times N_{rep}\)
The corresponding simulator outputs are denoted as \(y^{1,1},y^{1,2},\ldots,y^{n,i},\ldots, y^{N_{unique},N_{rep}}\).

A replicated design allows to pre-average the corresponding \(y\)-values, \(\bar{y}^n := \frac{1}{N_{rep}} \sum_{i=1}^{N_{rep}} y^{n,i}\), and then calling the regression module on the reduced dataset \((x^{1:N_{unique}},\bar{y}^{1:N_{unique}})\). Replication with pre-averaging offers a simple way of reducing the variance of observations. This is often desirable because the pathwise rewards tend to be highly volatile especially over longer periods of time, and many functional approximators struggle under low signal-to-noise settings. With high degree of replication, one can view $\bar{y}^n$ as almost deterministic, so that regression effectively reduces to interpolation.

\rev{{\em Remark:} As mentioned, since the entire algorithm proceeds step by step, $\cD$ can depend on $k$. Similarly, one can straightforwardly incorporate all types of time-dependency in the dynamics, costs, etc. Such a generalization is conceptually trivial and requires only careful encoding of further $k$-dependent objects.}

\subsection{Illustration}

To illustrate the overall workflow of solving an optimal impulse problem, I present a few brief code snippets. These utilize the \texttt{mlOSP} constructs and can be directly reproduced by any reader who installs the package. Consider impulsing a 1-D Geometric Brownian Motion (GBM) process with uncontrolled dynamics
\[ dX^\emptyset_t = \mu X^\emptyset_t dt + \sigma X^\emptyset_t dW_t, \qquad X^\emptyset_0 = x_0,\]
with scalar parameters \(\mu,\sigma,x_0\). Thus, \( (X^\emptyset_t)\) can be simulated exactly by sampling from the respective log-normal distribution.
The running payoff is of concave power-type $\pi(x) = x^\gamma/\gamma$, $0< \gamma < 1$, and the intervention costs are linear $\kappa(x,z) = c_0 \cdot z + c_1$. This setup is motivated by Federico et al.~\cite{Federico19} who considered irreversible investment with fixed adjustment costs. The state process $(X_t)$ represents an economic indicator, such as the production capacity of a firm which drives the revenue rate $\pi(X_t)$.

As discussed above, linear impulse costs yield an optimal strategy of $(s,S)$ type: intervene as soon as $(X_t)$ goes below $s$ and bring it back up to $S>s$.
Thanks to the linearity of GBM and $\kappa(x,\cdot)$, and the power-form of $\pi(x)$ the infinite-horizon problem is known to have an explicit solution
\begin{align}
  \tilde{v}(x) = B x^m + C x^\gamma/\gamma, & \qquad s = \left( \frac{ c_0 (m-1)}{C(m-\gamma)} \right)^\frac{1}{\gamma-1}, \\ \label{federico-2}
  m = \left( \frac{1}{2} - \frac{\mu}{\sigma^2} \right) - \sqrt{ \left( \frac{1}{2} - \frac{\mu}{\sigma^2} \right)^2 + \frac{2 r}{\sigma^2} }, &\qquad
  C = \frac{1}{ r - \mu \gamma + 0.5 \gamma (1-\gamma)\sigma^2 }, \qquad B = \frac{C(1-\gamma)}{m(m-1)} s^{\gamma-m}.
\end{align}
Thus, with infinite horizon the controlled $(X_t)$ will be a time-stationary renewal process and undergo a cyclical behavior with renewal times
$\tau_{n+1} = \tau_n + \inf \{t > 0 : X^{\tau_n,S}_{t} \le s \}$ and $z_{n+1} = S-s$.

To implement the above instance, one starts by defining the \texttt{model}, which is a list of (a) parameters that determine the dynamics of \((X^\emptyset_t)\) in \eqref{eq:sde}; (b) the running payoff function \(\pi(x)\); (c) the impulse function $\kappa(x,z)$ and (d) the tuning parameters determining the regression surrogate specification. In the example I take $r=0.08, \mu = -0.07, \sigma =0.25, c_0 = -1, c_1 = -10$ and square-root running reward $\gamma = 0.5$ which yields the time-stationary solution thresholds $s=8.749,S=56.99$. I  use a finite horizon of $T=10$ with $\Delta t = 0.1$, i.e.~$K=100$ time-steps.

\begin{Shaded}
\begin{Highlighting}[]
\NormalTok{modelFRT <-}\StringTok{ }\KeywordTok{list}\NormalTok{(}\DataTypeTok{dim=}\DecValTok{1}\NormalTok{, }
                   \DataTypeTok{sim.func=}\NormalTok{sim.gbm, }
                   \DataTypeTok{r=}\FloatTok{0.08}\NormalTok{,    }\CommentTok{# discount factor}
                   \DataTypeTok{div=}\FloatTok{0.15}\NormalTok{,  }\CommentTok{# drift is mu=-0.07 }
                   \DataTypeTok{sigma=}\FloatTok{0.25}\NormalTok{, }\CommentTok{# volatility}
                   \DataTypeTok{x0=}\DecValTok{50}\NormalTok{,    }\CommentTok{# initial state}
                   \DataTypeTok{impulse.fixed.cost =} \DecValTok{10}\NormalTok{, }\CommentTok{# fixed }
                   \DataTypeTok{impulse.cost.linear =} \DecValTok{1}\NormalTok{, }\CommentTok{# linear }
                   \DataTypeTok{impulse.func =}\NormalTok{ lin.impulse,}
                   \DataTypeTok{imp.type =} \StringTok{"exchrate"}\NormalTok{,}
                   \DataTypeTok{gamma =} \FloatTok{0.5}\NormalTok{,}
                   \DataTypeTok{running.func =} \ControlFlowTok{function}\NormalTok{(x)(}\DecValTok{2}\OperatorTok{*}\KeywordTok{sqrt}\NormalTok{(x)), }\CommentTok{# cont profit rate pi}
                   \DataTypeTok{T=}\DecValTok{20}\NormalTok{,   }\CommentTok{# horizon}
                   \DataTypeTok{dt=}\FloatTok{0.2}\NormalTok{,  }\CommentTok{# time step; 100 steps total}
                   \DataTypeTok{pilot.nsims=}\DecValTok{0}\NormalTok{,}
                   \DataTypeTok{batch.nrep =} \DecValTok{40}\NormalTok{,  }\CommentTok{# replicates for each unique input}
                   \DataTypeTok{N =} \DecValTok{600}\NormalTok{,  }\CommentTok{# N_unique training locations}
\NormalTok{              )}
\DataTypeTok{input.dom =} \KeywordTok{c}\NormalTok{(}\KeywordTok{seq}\NormalTok{(}\DecValTok{1}\NormalTok{,}\DecValTok{18}\NormalTok{,}\DataTypeTok{length=}\FloatTok{350}\NormalTok{), }\KeywordTok{seq}\NormalTok{(}\FloatTok{18.2}\NormalTok{,}\DecValTok{90}\NormalTok{,}\DataTypeTok{length=}\FloatTok{250}\NormalTok{))}
\end{Highlighting}
\end{Shaded}

Next, we must choose a solver scheme, namely specifying the surrogate type and the training sets. In \texttt{mlOSP}, the solver implementing Algorithm \ref{alg:1} is named \texttt{osp.impulse.control} and comes with a \texttt{method} field that controls the surrogate, and \texttt{input.domain} that controls the simulation designs. The code  below utilizes a \rev{(cross-validated) spline surrogate that automatically sets the number of knots}. For the \texttt{input.domain} I utilize a non-uniform lattice that is dense for $x \in [1,18]$ (region of the action set) and less so for $x \in [18,90]$. I also employ replication, with each input replicated \texttt{batch.nrep=40} times.

For the terminal condition I take the expected value of future running rewards given the current state and no more impulses, i.e.~$\widehat{Q}(K,x) = \phi(x) = \E_x \left[ \int_0^\infty e^{-r t}\pi(X^\emptyset_t) dt \right] = C x^\gamma/\gamma$ where $C$ is from \eqref{federico-2}. This is interpreted as $T$ being the horizon for actions, thereafter $X$ evolves endogenously without any further controls.

\begin{Shaded}
\begin{Highlighting}[]
\NormalTok{spl.solver <-}\StringTok{ }\KeywordTok{osp.impulse.control}\NormalTok{(modelFRT,}
                          \DataTypeTok{input.domain =}\NormalTok{ input.dom,}\DataTypeTok{method=}\StringTok{"cvspline"}\NormalTok{)}
\end{Highlighting}
\end{Shaded}

Note that no output is printed: the produced object \texttt{spl.solver} contains an array of 99 (one for each time step, except at maturity) fitted smoothing spline surrogates for $\widehat{Q}(k,\cdot)$. Technically, \texttt{spl.solver} is a list that has a few other diagnostics beyond the collection of the \texttt{smooth.spline} objects.

\begin{figure}

{\centering \begin{tabular}{ccc}\includegraphics[width=0.32\textwidth,trim=0.2in 0.2in 0in 0.2in]{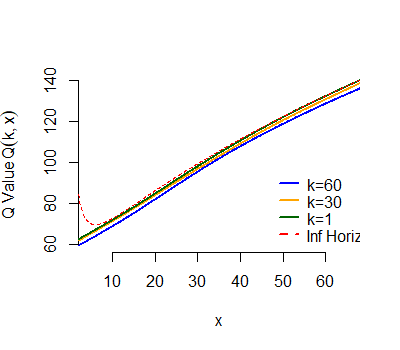} &
\includegraphics[width=0.32\textwidth,trim=0in 0.2in 0.2in 0.4in]{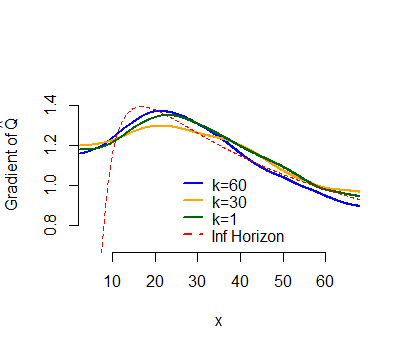} & \includegraphics[width=0.32\textwidth,trim=0in 0.2in 0in 0.4in]{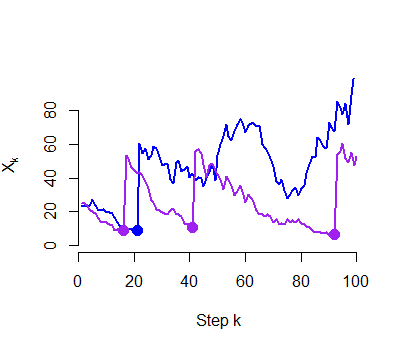}
\end{tabular}

}

\caption{\label{fig:1d}A 1D capacity expansion instance. Left: Q-value based on a Smoothing Spline emulator at $k \in \{1, 30, 60\}$. Middle: corresponding gradient $\partial_x \widehat{Q}(k,x)$. The target level $S^*_k$ is the threshold where $\partial_x \widehat{Q}(k,S^*_k) = 1$. Right: two resulting controlled paths of  $X^{\widehat{\mY}}$. Dots indicate the times $\tau^{n'}_k$ of impulses.
}\label{fig:Fitted-GP}
\end{figure}

Figure \ref{fig:1d} visualizes the fitted $\widehat{Q}$ from three time-steps. To do so, we simply predict the Q-value object over a collection of test locations. In the Figure, this is done for $k=1,30,60$. The middle panel shows the gradient $\partial_x \widehat{Q}(k,\cdot)$, obtained by finite-differencing, at the same time steps $k$.

To better understand  the resulting strategy $\widehat{\mY}_{0:K}$ we build an independent database of forward controlled paths. This is done via the \texttt{forward.impulse.policy} command that  evaluates \eqref{eq:tildeV} and also records all the associated actions (impulse amounts and times).
The right panel of Figure \ref{fig:1d} shows two different controlled forward paths based on the computed $\widehat{\mY}_{0:K}$. On the blue path there are 3 impulse times $\tau_n$; on the purple one only two. One can clearly see the $(s,S)$ policy where $(X^{\widehat{\mY}}_t)$ is impulsed whenever it gets too low and is then brought up to about $S^*_k \simeq 60$ which is the target level.

\begin{figure}
\centering
  \includegraphics[height=2.2in,trim=0.2in 0.3in 0.2in 0.3in]{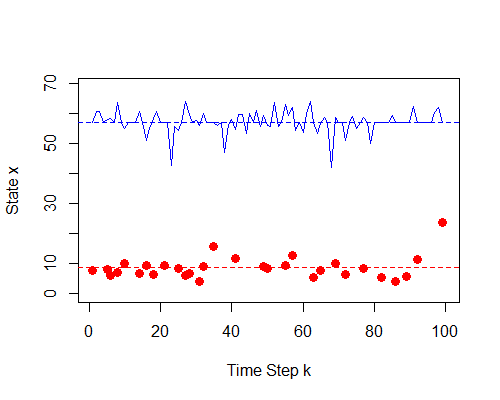}
  \caption{Threshold boundary $s^*_k$ (dots towards the bottom) and threshold target levels $S^*_k$ (line towards the top) for the irreversible investment case study. \label{fig:sS}}
\end{figure}

Below we also show code to plot the impulse strategy, namely the impulse boundary $s^*_k$ and the impulse target levels $S^*_k$, displayed in Figure \ref{fig:sS}. Note that the shown $s^*_k$ is based on the forward paths, so at some times $k$ there is no recorded $s^*_k$ since none of the forward paths were impulsed at that specific $k$ (impulses are not so frequent). The Figure shows the time-stationary $s,S$ values that confirm the good approximation by the present solver. One can also observe the time-dependence which manifests itself through the agent being impatient as problem horizon
is approached. As a result, impulses are applied sooner (lower timing value, i.e. lower opportunity cost of acting) and we see a ``boundary layer" as $k \to K$. The slight fluctuations observed in $s^*_k, S^*_k$ are due to Monte Carlo-driven approximation errors and can be decreased with larger training sets.

\begin{Shaded}
\begin{Highlighting}[]
\NormalTok{S.target <-}\StringTok{ }\KeywordTok{rep}\NormalTok{(}\DecValTok{0}\NormalTok{,}\DecValTok{100}\NormalTok{)}

\ControlFlowTok{for}\NormalTok{ (j }\ControlFlowTok{in} \DecValTok{1}\OperatorTok{:}\DecValTok{99}\NormalTok{)}
\NormalTok{  S.target[j] <-}\StringTok{ }\KeywordTok{lin.impulse}\NormalTok{(}\KeywordTok{seq}\NormalTok{(}\DecValTok{1}\NormalTok{,}\DecValTok{10}\NormalTok{,}\DataTypeTok{by=}\DecValTok{1}\NormalTok{),modelFRT,}
      \NormalTok{spl.solver}\OperatorTok{$}\NormalTok{fit[[j]],}\DataTypeTok{ext=}\OtherTok{TRUE}\NormalTok{)}\OperatorTok{$}\NormalTok{imp.target[}\DecValTok{1}\NormalTok{]}

\CommentTok{# forward paths}
\NormalTok{fi <-}\StringTok{ }\KeywordTok{forward.impulse.policy}\NormalTok{(}\KeywordTok{array}\NormalTok{(}\DecValTok{10}\NormalTok{,}\DataTypeTok{dim=}\KeywordTok{c}\NormalTok{(}\DecValTok{10000}\NormalTok{,}\DecValTok{1}\NormalTok{)) , }\DecValTok{100}\NormalTok{, spl.solver}\OperatorTok{$}\NormalTok{fit,modelFRT)}
\CommentTok{# these are the s-values}
\KeywordTok{plot}\NormalTok{(fi}\OperatorTok{$}\NormalTok{bnd, }\DataTypeTok{xlab=}\StringTok{'Time Step k'}\NormalTok{, }\DataTypeTok{ylab=}\StringTok{'State x'}\NormalTok{, }\DataTypeTok{pch=}\DecValTok{19}\NormalTok{,}\DataTypeTok{cex=}\FloatTok{1.2}\NormalTok{, }\DataTypeTok{col=}\StringTok{"red"}\NormalTok{, }\DataTypeTok{ylim=}\KeywordTok{c}\NormalTok{(}\DecValTok{0}\NormalTok{,}\DecValTok{65}\NormalTok{))}
\KeywordTok{points}\NormalTok{(S.target[}\DecValTok{1}\OperatorTok{:}\DecValTok{99}\NormalTok{], }\DataTypeTok{cex=}\FloatTok{1.2}\NormalTok{, }\DataTypeTok{col=}\StringTok{"blue"}\NormalTok{,}\DataTypeTok{pch=}\DecValTok{19}\NormalTok{)}
\KeywordTok{abline}\NormalTok{(}\DataTypeTok{h=}\KeywordTok{c}\NormalTok{(}\FloatTok{56.99}\NormalTok{,}\FloatTok{8.749}\NormalTok{),}\DataTypeTok{lty=}\DecValTok{2}\NormalTok{,}\DataTypeTok{col=}\KeywordTok{c}\NormalTok{(}\StringTok{"blue"}\NormalTok{,}\StringTok{"red"}\NormalTok{))} \CommentTok{# solution of the inf-horizon}
\end{Highlighting}
\end{Shaded}

\section{Case Studies}\label{sec:results}
In this section I present two more case studies that showcase other problem settings and further features available in \texttt{mlOSP}.

\subsection{Faustmann Problem of Forest Rotation}\label{sec:forest}

For the next example of an impulse control problem amenable to Algorithm \ref{alg:1}, I take up the problem of forest management as nicely summarized and analyzed by Alvarez and co-authors \cite{alvarez2007optimal,alvarez2008optimal}. Let $X_t$ represent the value of forest stand, i.e.~the economic value of existing timber. Forest growth is modeled as a stochastic process: timber increase is uncertain and fluctuates due to weather, precipitation and other environmental effects. Droughts or insect infestations might reduce forest stand, justifying the use of stochastic differential equations for modeling $(X_t)$.

The controller aims to maximize total profit from cutting down and selling timber on a given time horizon $T$. This is achieved through carrying out a sequence of so-called forest rotations. At each rotation, the forest stand is cut down and sold. The number of rotations is stochastic and up to the forest manager. The horizon $T$ represents the lease term for the timberland, measured in years. The objective functional is then
\begin{align}
  \E \left[ \sum_{n : \tau_n <T} e^{-r \tau_n} \kappa(X_{\tau_n-}, z_n)  \right],
\end{align}
where the payoff function $\kappa$ captures the value of selling $z_n$ timber, subject to the cutting costs, and $r$ is the intertemporal discount rate. Above we assume zero salvage value at $T$, $\phi(X_T) \equiv 0$.  As might be expected, the timber will be cut at some time-dependent threshold \rev{$S^*_k$}.

In the classic formulation, the problem is stated on infinite horizon, the dynamics of $X_t$ are linear and time-homogeneous, and the post-rotation level $X_{\tau_k} \equiv \underline{x}$ is pre-specified. This offers an explicit solution, see \cite{alvarez2007optimal}, who showed that the action region is $\mathfrak{S}_k = \{ x > S^*\}$ where $S^*$ is the maximizer of a certain nonlinear equation. The finite-horizon version is analyzed in \cite{belak2017general} and I reproduce their example where $(X^\emptyset_t)$ is a standard arithmetic Brownian Motion, $r=0.1$, and there is a fixed cost for each rotation, $\kappa(x,z) = (z-1)_+$. Thus,  the forest is always cut down to nominal level zero $\underline{x} = 0$. This means that $z_n = X_{\tau_n-}$. The reference impulse boundary is $S^* = 1.84$.

I take a horizon of $T=5$ years and time steps of $\Delta t = 0.1$, yielding $K=50$ periods. For the regression, I utilize a Gaussian Process emulator with the squared-exponential kernel \eqref{eqn:se-kernel} and hyperparameters fitted via Maximum Likelihood Estimation, as done in the \texttt{DiceKriging} package. I then use the exact surrogate gradient based on \eqref{eq:grad-mean} to efficiently find $S^*_k$ in \eqref{eq:Z-star}.
 For the training simulation design, the particular structure of this formulation implies that it is important to obtain an accurate estimate of $\widehat{Q}(k,0)$. To this end, I consider training in a Monte Carlo forest like fashion, employing a high degree of replication so that each unique input will have multiple forward paths emanating from it. Namely, I take 100 unique inputs on a lattice between $[-0.25,2.5]$, each replicated 100 times, for a total of 10,000 forward training paths.

\begin{figure}
\includegraphics[width=.47\textwidth]{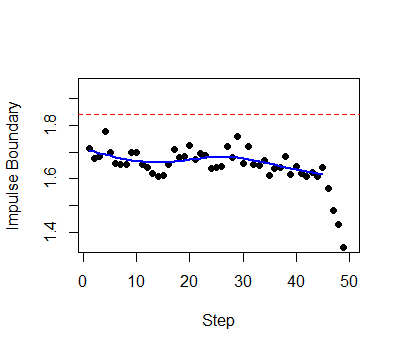}
\includegraphics[width=.47\textwidth]{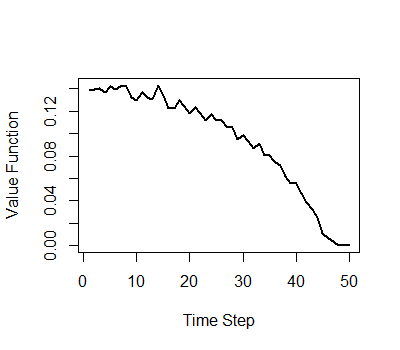}
\caption{Left: estimated impulse boundary $\hat{S}^*_k$ as a function of time step $k$, infinite horizon threshold $S^*$ is shown as a dashed line. To smooth out minor numerical artifacts we also display a smoothed estimate of $S^*$  as a blue curve. Right: value function $\widehat{V}(k,0)$ at zero. \label{fig:faustmann}}
\end{figure}

The left panel of Figure \ref{fig:faustmann} effectively reproduces Figure 1 in Belak et al.~\cite{belak2017general}; in contrast to that paper where the impulse boundary is obtained from a solution of an integral equation and requires specific assumptions on the impulse cost function and dynamics of $(X_t)$, my method is completely generic and can be trivially re-solved if any of the ingredients were to change. In Figure \ref{fig:faustmann} we can clearly observe the effect of the terminal condition, where the manager will cut down even a bit of forest ahead of the deadline $T$ that would give him no profit whatsoever. One also notes that the estimated impulse boundary is below that of the infinite-horizon problem. This arises due to (i) restricting actions to take place in $\cT_k$, here with separation $\Delta t = 0.2$, this is known to induce the manager to act sooner; (ii) the finite horizon that remains non-negligible with $T=5$, manifested by the impulse boundary slowly moving up as $k$ decreases. The right panel of Figure \ref{fig:faustmann} shows the estimated value $\widehat{V}(k,0)$ which also displays time-dependence and indicates convergence (i.e.~approaching time-stationary) with about 5 years until maturity.

\subsection{Two Dimensional Capacity Expansion}\label{sec:capexpand}

In the 2-dimensional version of capacity expansion, the state processes are the production price $(P_t)$ and the current capacity $C_t$. The price is exogenous and stochastic, while the capacity is fully endogenous and deterministic. We refer to \cite{bensoussan2019sequential} and \cite{guthrie2012uncertainty} who provided explicit solutions (up to solving an integral equation) in some special cases.

The price follows Geometric Brownian motion
$$
dP_t = \mu P_t dt + \sigma P_t dW_t
$$
and the capacity undergoes deterministic exponential decay/aging with rate $\delta$:
$$
C_t = e^{-\delta t} C_0  + \sum _k z_k e^{-\delta(t-\tau_k)}
$$
with profit function $\pi(P_t, C_t) = P_t C_t^\alpha$ for $\alpha < 1$, representing decreasing efficiency of adding more capacity. We identify above with a two dimensional state $X_t \in \R_+^2$ where the impulses $z \in \mathbb{R}_+$ only affect the second coordinate: an impulse of size $z$ leads to $P_{\tau_n} = P_{\tau_n-}$ and $C_{\tau_n} = C_{\tau_n-} + z$. Obvious generalizations to handle the two coordinates are straightforward to handle in code and are effectively abstracted away by the package as far as the user is concerned.

Following \cite{guthrie2012uncertainty} I consider concave investment costs $\kappa( (p,c),z) = z^\beta$ with $\beta < 1$ and $\beta > \alpha$. The concavity of $\kappa$ encourages making large investments. The quantity $Y_t := P_t C_t^{\alpha-\beta}$ can be interpreted as the firm's return on assets (ROA) and affords dimension reduction in the case of log-linear dynamics as above. Indeed \cite{guthrie2012uncertainty} shows that with these choices and infinite horizon, $V(p,c) = c^\beta v( c^{\alpha-\beta}p)$ where $v(y) = B_0 y + B_1 y^\gamma$ for some explicit constants $B_0,B_1,\gamma$ and the optimal impulses are of $(s,S)$-type in  $Y_t$, rather than $P_t, C_t$ separately.

Guthrie \cite{guthrie2012uncertainty} considers the parameter values $r=0.04,  \mu = 0, \sigma = 0.08, \delta = 0.1, \beta = 0.95, \alpha = 0.5$ which gives ROA threshold $y_0 = 0.224$. Moreover, the respective optimal impulse is to increase capacity by 178; this happens on average once every 11 years. In this setup, the problem is strongly non-stationary: $(P_t)$ is autonomous and can grow without bound (since $\mu > 0$) while $C_t$ is impulsed upwards. Consequently, one must select time-dependent simulation designs $\cD_k$ lest the forward paths end up extrapolating, rather than interpolating $\widehat{Q}$.

\begin{figure}

\includegraphics[width=0.37\textwidth,trim=0in 0in 0.2in 0in,clip=TRUE]{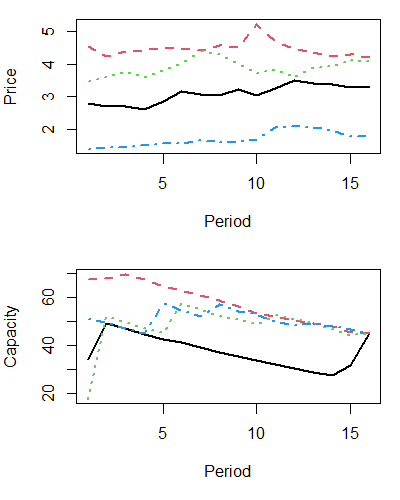}
\includegraphics[width=0.62\textwidth,trim=1.1in 0.5in 0.2in 0.8in, clip=TRUE]{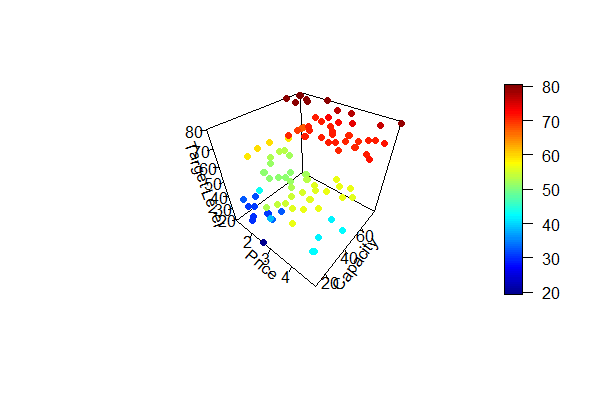}
\caption{Two dimensional finite horizon impulse control problem inspired by \cite{guthrie2012uncertainty}. Left: 4 controlled trajectories of $X_t=(P_t, C_t)$. Capacity $C_t$ decays exponentially without impulses. Right: impulse target $(p,c) \mapsto c +\widehat{\mY}_k(p,c)$ for a representative time step $k$. \label{fig:guthrie}}
\end{figure}

For the terminal condition we set $\phi( x) = \E [ \int_0^\infty e^{-r s}\pi(X_s) ds | X_0 = (p,c)] = p c^\alpha \frac{1}{r-\mu}$. There is no simple way to summarize the resulting solution; Figure \ref{fig:guthrie} displays a few optimally controlled paths, as well as the impulse target map (namely the $\arg\max$ of $\widehat{M}(k,\cdot)$) on the training set $\cD_k$, here chosen to be a Sobol low-discrepancy sequence. We note that with the nonlinear impulse costs, the impulse target depends nontrivially on both coordinates, increasing both in price $p$ and in current capacity $c$.

  \bibliography{rmcBook,impulseControl} 

\end{document}